\documentclass[12pt, draft]{iopart}
\usepackage{iopams}

\begin{document}

\newcommand{\eqref}[1]{(\ref{#1})}
\newcommand{\operatorname}[1]{\mbox{#1}}
\newcommand\half{{\textstyle \frac12}}

\renewcommand{\d}{\operatorname{d}}
\newcommand{\E}{\operatorname{E}}
\newcommand{\trace}{\operatorname{trace}}
\newcommand{\openR}{{\mathbb R}}
\newcommand{\openD}{{\mathbb D}}
\newcommand{\bra}[1]{\left \langle #1 \right |}
\newcommand{\ket}[1]{\left | #1 \right \rangle}
\newcommand{\braket}[2]{\left \langle #1 \mid  #2 \right \rangle}
\newcommand{\ketbra}[2]{\ket{#1}\bra{#2}}
\newcommand{\braopket}[3]{\left \langle #1 \mid  #2 \mid  #3\right \rangle}
\newcommand{\braopopket}[4]{\left \langle #1 \mid  #2 \mid  #3 \mid #4 \right \rangle}

\title{Fisher information in quantum statistics}

\author{O. E. Barndorff-Nielsen}

\address{MaPhySto, Aarhus University, 8000 Aarhus-C, Denmark}

\author{R. D. Gill}

\address{Mathematical Institute, University of Utrecht, Box 80010,
                 3508 TA Utrecht, NL; and
           EURANDOM, Box 513, 5600 MB Eindhoven, Netherlands}

\date{today}

\begin{abstract}

Braunstein and Caves (1994) proposed to use Helstrom's
{\em quantum information\/} number to define, meaningfully,
a metric on the set of all possible states of a given quantum system.
They showed that the quantum information is nothing else
than the maximal Fisher information in a measurement of the
quantum system, maximized over all possible measurements.
Combining this fact with classical statistical results, they argued
that the quantum information determines the asymptotically optimal
rate at which neighbouring states on some smooth curve can be
distinguished, based on arbitrary measurements on $n$ identical
copies of the given quantum system.

We show that the measurement which maximizes the Fisher
information typically depends on the true, unknown, state of
the quantum system. We close the resulting loophole in the argument
by showing that one can still achieve the same, optimal, rate of
distinguishability, by a two stage adaptive measurement procedure.

When we consider states lying not on a smooth curve,
but on a manifold of higher dimension, the situation becomes much
more complex. We show that the notion of
``distinguishability of close-by states'' depends strongly on
the measurement resources one allows oneself, and on a further
specification of the task at hand. The quantum information
matrix no longer seems to play a central role.

\end{abstract}


\maketitle

\section{Introduction}\label{intro}

Braunstein and Caves (1994) have clarified the relation between
the classical Fisher expected information number $i(\theta)$, for the
unknown parameter $\theta$ of a probability distribution $p(x;\theta)$,
and the analogous concept of expected quantum information $I(\theta)$
for a quantum system in state $\rho =\rho(\theta)$ on some Hilbert space.
They showed that $I(\theta)$ is the maximal Fisher information $i(\theta;M)$
in the distribution of the outcome of a measurement $M$, over all measurements
of the state. Thereby they supplied a new proof of
Helstrom's (1967) quantum Cram\'er-Rao bound:
no unbiased estimator of $\theta$, based on any
measurement, has variance smaller than $I(\theta)^{-1}$.
Recall that the classical bound states that no unbiased estimator of $\theta$
based on the outcome of the measurement $M$ has variance smaller than
$i(\theta;M)^{-1}$.

For $n$ identical copies of a quantum system, and for $n$
independent and identically distributed observations
from a probability distribution,
quantum and Fisher information are both $n$ times the corresponding quantities
for $n=1$. By classical statistical theory, the quantum bound is therefore
{\em asymptotically\/} achieved, as $n\to\infty$, by the maximum likelihood
estimator of $\theta$ based on the outcomes of the measurement maximizing
the Fisher information for $n=1$, applied to each of $n$ copies of the quantum
system separately.

In the present paper, we analyse the conditions for equality of the quantum
and Fisher information. We show that in general there does not exist a
measurement $M$ such that $i(\theta;M)=I(\theta)$ for all $\theta$
simultaneously, studying the pure state, spin-half case in detail.
In that case the model describes a curve on the surface of the unit
sphere, specifying the direction of the spin as a function of $\theta$.
We show that one has uniform attainability if and only if the
curve is a segment of a great circle. We show how (in general)
adaptive measurements still allow one to asymptotically achieve the quantum
information bound for a scalar parameter, though not in the vector case,
where the
picture is rather complicated and the quantum information matrix inadequate
to describe what is possible.

In Section 2 of the paper we recapitulate some of the theory of
classical and quantum information. Next, in Section 3, we specialize the
conditions for attainability of the information bound, first to
pure states, then further to spin-half models.
Unless the model specifies a great circle, no
measurement achieves the bound uniformly in the parameter $\theta$.
In Section 4 we explore the consequences of this result. We show that
one can in effect achieve $i(\theta;M_n) \approx n I(\theta)$ for all
$\theta$ simultaneously, when we measure
$n$ identical copies of the quantum system in one joint measurement $M_n$.
This result gives support to Braunstein and Cave's interpretation
of the quantum information number $I(\theta)$ as a measure of statistical
distinguishability between neighbouring quantum states.
Finally we turn to the case when the parameter is a vector. Both quantum
and classical information numbers have matrix generalizations, and
inequality between them still holds, in the sense of positive semi-definite
matrices. The inequality is sharp but however no longer attainable.
For a completely unknown spin-half pure state we show that the optimal rate at
which one can distinguish between different states does not follow from the
quantum information matrix in the way one would expect from analogy with
classical Fisher information. Moreover it depends on some weighting
of the different aspects of the states which one wants to distinguish.
Major open problems remain, and the role of the quantum information does not
appear to be primary.

A preliminary version of this paper appeared as Barndorff-Nielsen
and Gill (1998).

\section{Expected classical and quantum information}\label{recap}

On a given Hilbert space, consider a quantum state (density operator)
$\rho=\rho(\theta)$, which depends on an unknown scalar parameter $\theta$.
Consider also a generalised measurement (operator-valued probability measure,
POVM) $M$ with outcomes in a measurable space $(\mathcal{X},\mathcal{A})$.
Thus the outcome of a measurement of $M$ on $\rho$
is a random variable $X$ taking values in
$\mathcal{X}$, such that for each measurable subset $A$ of
$\mathcal{X}$, i.e., for each $A\in\mathcal{A}$,
we have $\Pr_\theta\{X\in A\}=\trace\,\rho(\theta)M(A)$.
Suppose that $M$ is dominated by a sigma-finite measure
$\mu$ on $(\mathcal{X},\mathcal{A})$, i.e., for each $A\in\mathcal{A}$,
\begin{equation*}
M(A) ~ = ~ \int_A m(x)\mu(\d x)
\end{equation*}
where the operator $m(x)$ is, for each $x$, nonnegative and selfadjoint,
and $\int_{\mathcal{X}}m(x)\mu(\d x)=\mathbf 1$. (This is no
restriction for finite dimensional quantum systems, for which one
can always take $\mu$ to be the measure defined by $\mu(A)=\trace\,M(A)$.)
Under the domination assumption, the outcome $X$ of a measurement of
$M$ on $\rho$ has probability density, with respect to $\mu$, given by
\begin{equation*}
p(x;\theta) ~ = ~ \trace \, \rho(\theta) m(x).
\end{equation*}
Under sufficent smoothness, the
expected {\em Fisher information\/} number, for $\theta$,
from this measurement, is defined by
\begin{equation*}
i(\theta;M) ~ = ~ \E_\theta \dot l(\theta)^2
 ~ = ~ \int_{\mathcal{X}}(\dot l(x;\theta) )^2 p(x;\theta )\mu(\d x)
\end{equation*}
where
\begin{equation*}
l(\theta) ~ = ~ l(X;\theta) ~ = ~ \log p(X;\theta)
\end{equation*}
is the {\em log likelihood\/} and
\begin{equation*}
\dot l(\theta) ~ = ~ \frac \partial {\partial \theta} l(X;\theta)
\end{equation*}
is the {\em score function\/} for $\theta$.

Now, let $\lambda=\lambda(\theta)$ denote the symmetric logarithmic
derivative of $\rho$ with respect to $\theta$, that is,
the self-adjoint operator given implicitly by
\begin{equation}
\dot\rho ~ = ~ \half (\rho\lambda +\lambda\rho).
\label{sld}
\end{equation}
We call $\lambda$ the {\em quantum score\/} for $\theta$. From the
relation $\trace\,\rho=1$ one finds, by differentiating, $\trace\,\rho\lambda=0$.
The expected {\em quantum information\/} number for $\theta$ is defined by
\begin{equation*}
I(\theta) ~ = ~ \trace \, \rho\lambda^2.
\end{equation*}
Note that this quantity is defined without reference to any
particular measurement $M$.

For future reference, we mention that when $\theta$ is a vector parameter,
the Fisher information matrix is defined in the obvious way, while
the quantum information matrix has $ij$-component
$\half\trace\,\rho(\lambda_i\lambda_j+\lambda_j\lambda_i)$ where
$\lambda_i(\theta)$ is the quantum score for $\theta_i$ keeping the
other components of $\theta$ fixed. For completeness, we mention that
there exist other generalizations of quantum information; see
Yuen and Lax (1973), Belavkin (1976) and the books Helstrom (1976),
Holevo (1980).

Define
\begin{equation}
{\mathcal{X}}_0  ~ = ~ {\mathcal{X}}_0(M,\theta) ~ = ~
\{x:p(x;\theta)=0\}
\end{equation}
and let ${\mathcal{X}}_+$ be its complement.
One can express the Fisher information $i(\theta;M)$ in terms of
the quantum score for $\rho$:
\begin{equation*}
i(\theta;M)~=~\int_{{\mathcal{X}}_+} p(x;\theta)^{-1}
\bigl( \Re \,\trace(\rho \lambda m(x) ) \bigr)^2 \mu(\d x) .
\end{equation*}
This follows on noting that
\begin{eqnarray*}
  \dot l(\theta)  ~ &= ~  p(x;\theta)^{-1}\trace \, \dot\rho m(x) \\
    &= ~ p(x;\theta)^{-1}\half\trace((\rho\lambda+\lambda\rho) m(x) ) \\
    &= ~ p(x;\theta)^{-1}\Re\,\trace(\rho\lambda m(x)).
\end{eqnarray*}

The usual proof of the quantum Cram\'er-Rao inequality,
see Helstrom (1976) or, for a more abstract and precise
version, Holevo (1982), follows closely the lines of the usual proof
of the classical bound:
write down the unbiasedness relation, differentiate under the
integral sign, and apply the Cauchy-Schwarz inequality;
Holevo (1982) gave a more rigorous proof on the same lines.
Braunstein and Caves (1994) noted how the quantum bound could be obtained
from the classical bound together with their new inequality
$i(\theta;M)\le I(\theta)$ for all measurements $M$.
The derivation is
a chain of three inequalities and therefore leads to a set of
three necessary and sufficient conditions for equality (though
they did not notice the third).
Before presenting the derivation we list the three ingredients.
For a given $x\in{\mathcal{X}}_+$, let $A=m(x)^{1/2} \rho^{1/2}$,
$B=m(x)^{1/2} \lambda \rho^{1/2}$, and $z=\trace(A^* B)$. The
the first inequality step uses the trivial
$(\Re(z))^2\le |z|^2$ with equality if and only if $\Im(z)=0$. The
second uses the Cauchy-Schwarz inequality
$| \trace(A^* B) |^2 \le \trace(A^* A) \trace(B^* B) $
with equality if and only if $\trace(A^* A) B= \trace(A^* B) A$.
The third inequality step,  $\trace( M({\mathcal{X}}_+)\lambda
\rho\lambda) \le \trace(\rho\lambda^2)$, follows from the fact that
$M({\mathcal{X}}_+)=\mathbf 1 - M({\mathcal{X}}_0)$ where
$M({\mathcal{X}}_0)\ge\mathbf 0$. The three ingredients are put
together as follows:
\begin{eqnarray}
  i(\theta;M)  ~ & = ~  \int_{{\mathcal{X}}_+}
        p(x;\theta)^{-1} ( \Re\,\trace(\rho\lambda m(x) ) )^2 \mu(\d x)
                                                               \nonumber \\
   & \le ~  \int_{{\mathcal{X}}_+} p(x;\theta)^{-1}
                       |\trace(\rho\lambda m(x))|^2  \mu(\d x)
                                                               \nonumber \\
   & = ~  \int_{{\mathcal{X}}_+} \biggl|
         \trace \bigl( \, ( m(x)^\half \rho^\half )^* \,
                  ( m(x)^\half \lambda \rho^\half) \, \bigr)
             \biggr|^2 (\trace(\rho m(x)))^{-1}\mu(\d x)       \nonumber  \\
   & \le ~  \int_{{\mathcal{X}}_+}
        \trace( m(x) \lambda \rho \lambda ) \mu(\d x)          \nonumber \\
   & = ~\trace(M({\mathcal{X}}_+)\lambda \rho \lambda )         \nonumber \\
   & \le  ~  \trace(\rho\lambda^2) ~ = ~     I(\theta)~ .
                                                             \label{QIineq}
\end{eqnarray}
With $A$, $B$ and $z=\trace(A^* B)$ as above (depending on $x$),
necessary and sufficient conditions for equality at the first
two inequality steps in \eqref{QIineq} together are equivalent to:
for $\mu(\d x)$ almost all $x$ in ${\mathcal{X}}_+$, $\trace(A^* B)$
is real and $A\propto_\openR B$, by which we mean $A=rB$ or $B=rA$
for some real number $r$. But if $A\propto_\openR B$ then
automatically $\trace(A^* B)$ is real. Thus we have equality in
\eqref{QIineq} if and only if the following two conditions are
satisfied: firstly, for $\mu(\d x)$ almost all $x$ in ${\mathcal{X}}_+$
\begin{equation}
m(x)^\half \lambda \rho^\half ~ \propto_\openR ~ m(x)^\half \rho^\half
\label{cond1}
\end{equation}
and secondly,
\begin{equation}
\trace(M({\mathcal{X}}_0)\lambda \rho \lambda ) = 0.
\label{cond2}
\end{equation}
Obviously a sufficient condition for \eqref{cond2}
is that $M({\mathcal{X}}_0)=\mathbf 0$,
and a sufficient condition for that is $p(x;\theta) > 0$ for all
$\mu$-almost all $x$.
Braunstein and Caves remark that a sufficient condition for \eqref{cond1}
is that each $m(x)$ is proportional to a projector onto an eigenspace of
$\lambda$. In particular, if the measurement $M$ is a simple (von Neumann)
measurement of the observable $\lambda$ then \eqref{cond1} is satisfied.
However this is not a necessary condition for attainability. Thus the
obvious fact that, in general, $\lambda(\theta)$ varies with $\theta$,
does not show that there are no measurements attaining the bound
\eqref{QIineq} for all $\theta$ simultaneously. We will do this by
a further study of condition \eqref{cond1} in a special case.

\section{Attainability of the quantum information bound}

In this section we concentrate on models for pure
states, $\rho=\ketbra\psi\psi$ where $\ket\psi=\ket{\psi(\theta)}$.
In this case, the quantum score can be computed explicitly
and the condition \eqref{cond1} simplifies.
Define the (unnormalized) state $\ket a=2\dot{\ket\psi}$.
Since $\rho^2=\rho$, we have $\dot\rho=\rho\dot\rho+\dot\rho\rho$.
The defining equation \eqref{sld} for the quantum score therefore tells us
that $\lambda=2\dot\rho=\ketbra a \psi +\ketbra \psi a$.
Now let $\ket 1=\ket\psi$ and let $\ket 2$ be a normalized orthogonal
state such that $\ket a$ is in the subspace spanned by $\ket 1$ and $\ket 2$;
write $\ket a = a_1\ket 1 + a_2 \ket 2$ where $a_1=\braket 1 a$ and
$a_2 =\braket 2 a $.  (Note that all these definitions are relative to
a given value of the parameter $\theta$.) We find that
\begin{equation*}
\rho^\half\lambda ~ = ~ \rho\lambda ~ = ~
    2 (\Re\, a_1) \ketbra 1 1 + \overline a_2 \ketbra 1 2
\end{equation*}
and condition \eqref{cond1} reduces to
\begin{equation*}
m(x)^\half ( 2 (\Re \, a_1) \ketbra 1 1 +  a_2 \ketbra 2 1 )
    ~ \propto_\openR ~ m(x)^\half \ketbra 1 1  .
\end{equation*}
This can be again simplified, resulting in the condition
\begin{equation*}
a_2 m(x)^\half  \ketbra 2 1
    ~ \propto_\openR ~ m(x)^\half \ketbra 1 1
\end{equation*}
or equivalently
\begin{equation}
a_2 m(x)^\half  \ket 2
    ~ \propto_\openR ~ m(x)^\half \ket 1 .
\label{condpure}
\end{equation}

For spin-half models, thus a Hilbert space of dimension $2$, a further
simplification occurs. One can take $\ket 2 ={\ket \psi}^\perp$, forming
an orthonormal basis (depending on $\theta$) with $\ket 1=\ket \psi$.
>From \eqref{condpure} it follows that if $m(x)$ satisfies
\eqref{cond1}, it must have less than full rank, and hence in the
two-dimensional case both it and its square root must be proportional
(with real constants of proportionality) to $\ketbra \xi \xi$ for
some normalized state $\ket \xi = \ket{\xi(x)}$. A minor rewriting yields
that \eqref{cond1} is equivalent to the statement:
for $p(x;\theta)\mu(\d x)$ almost all $x$, $m(x)$ is proportional to a
one-dimensional projector $\ketbra {\xi(x)} {\xi(x)}$ satisfying
\begin{equation}
\braket \xi 2 \braket 2 a ~ \propto_\openR ~ \braket \xi 1  .
\label{condpurehalf}
\end{equation}

We show that this algebraic condition has a simple geometric
interpretation. First note that from the definition of $\ket a$
and the fact that  $\braket\psi\psi=1$ for
all $\theta$, it follows that $2\Re\braket a 1 = \braket a 1 + \braket 1
a = 0$, hence $\braket a 1$ is purely imaginary. By multiplying
$\ket 2$ by a suitable phase factor, one can arrange that
$\braket a 2 $ is real and \eqref{condpurehalf} becomes
$\braket \xi 2\propto_\openR\braket \xi 1$. Note that $2\dot\rho=\ketbra 1 a + \ketbra a 1$.
It follows that $\braopket 1 {\dot\rho} 1 = 0 = \braopket 2
{\dot\rho} 2 $, while $\braopket 1 {\dot\rho} 2 $ is real. Hence
$2\dot\rho= r(\ketbra 1 2 + \ketbra 2 1 )$ for some real number $r$.
Let $\sigma_x=\ketbra1 2 +\ketbra 2 1$, $\sigma_y=-i\ketbra 1 2 +i
\ketbra 2 1$, $\sigma_z=\ketbra 1 2+\ketbra 2 1$ be the Pauli spin matrices
with respect to the basis $\ket 1, \ket 2$. In this basis
$\rho=\half(\mathbf 1 +\sigma_z)$ and $2\dot\rho=r\sigma_x$.
We can write $\ketbra\xi\xi=\half(\mathbf 1+\vec\alpha\cdot\vec\sigma)$
where $\vec\alpha=(\alpha_x,\alpha_y,\alpha_z)$ is a unit vector
in $\openR^3$. Note that $\braket 2 \xi \braket \xi 1 = \alpha_x+i
\alpha_y$. Therefore \eqref{condpurehalf} holds if and only if
$\alpha_y=0$.

With respect to an arbitrary {\em fixed\/} basis we can write
$\rho(\theta)=\half(\mathbf 1 +\vec u(\theta)\cdot \vec\sigma)$
and $\lambda(\theta)=2\dot\rho(\theta)=\dot{\vec
u(\theta)}\cdot\vec\sigma=
r(\theta)\vec v(\theta)\cdot\vec\sigma$ where
$\vec u(\theta)$ and $\vec v(\theta)$ are orthogonal unit vectors,
$r(\theta)=\|\dot{\vec u(\theta)}\|$ is real and nonnegative,
and $\vec\sigma$ is the vector
of the Pauli spin matrices with respect to
the fixed basis. Using the familiar relations
$\sigma_x^2={\bf 1}$, $\sigma_x\sigma_y=-\sigma_y\sigma_x=i\sigma_z$,
and their cyclic permutations,
and the fact that the spin matrices are traceless,
one finds that the quantum information $I(\theta)=r(\theta)^2$.
The direction
$\vec v(\theta)$ is uniquely defined if $r(\theta)>0$ and from now
on we assume this is true for all $\theta$.
Then \eqref{cond1} is satisfied if and only if for $p(x;\theta)\mu(\d
x)$ almost all $x$,
$m(x)$ is proportional to a projector for a spin direction in the
plane spanned by $\vec u(\theta)$ and $\dot{\vec u(\theta)}$. In particular,
{\em any\/} simple (von Neumann) measurement of spin in a direction
in this plane attains equality in \eqref{QIineq} if both outcomes have
positive probability.

Let $\mathcal C (\theta)$ denote the great circle on the unit sphere
formed by the intersection of the sphere with the plane $\mathcal P (\theta)$
spanned by $\vec u (\theta)$  and $\vec v (\theta)$,
let $\vec n(\theta)$ be the normal unit vector
to this plane. We suppose that these objects vary smoothly with
$\theta$. Recall that $\vec u(\theta)$ moves in the direction $\vec
v(\theta)$.  Now as $\theta$ varies, either $\mathcal P (\theta)$,
$\mathcal C (\theta)$, and $\vec n(\theta)$ are all fixed or they all vary.
In particular, the intersection over all $\theta$ of the planes
$\mathcal P (\theta)$ is either a fixed plane $\mathcal P$
(equal to $\mathcal
P(\theta)$ for all $\theta$), or a fixed straight line $\mathcal L$,
or the origin $\vec 0$.
In the second case the normal $\vec n(\theta)$ and the circle
$\mathcal C (\theta)$ must be rotating about the fixed line $\mathcal L$.
The rotation must have nonzero speed for values of $\theta$
in a set of positive measure. If we assume that the model
is identified, so that $\rho(\theta)$ is a one-to-one
function of $\theta$, then at most for two values of $\theta$
can the true spin $\vec u(\theta)$ lie in $\mathcal L$.
So there exists a $\theta$ for which
the rotation has nonzero speed and $\vec u(\theta)$
does not lie in the axis of rotation. But then at this
point the derivative of $\vec u(\theta)$ must have a nonzero
component in the direction orthogonal to the plane
$\mathcal P (\theta)$, which is a contradiction.

So we only have two possiblities:
either the plane $\mathcal P(\theta)$ is fixed and
$\vec u(\theta)$ is moving on the great circle
$\mathcal C$ in the plane, or $\vec u(\theta)$ moves on some other
curve and the intersection of all $\mathcal P(\theta)$
contains only the origin.
In the first case any measurement with all components proportional to
projectors of directions in this plane, and with $p(x;\theta)>0$
for all $x$ and $\theta$, achieves the inequality \eqref{QIineq}
uniformly in $\theta$. Conversely, under the positivity of $p(x;\theta)$,
only such measurements uniformly achieve the bound.
In the second case a measurement which uniformly achieves
the bound would have to have all $m(x)$ equal to
zero, which is impossible.  Thus there
is no uniformly attaining measurement in this case.

For example, consider a spin-half particle in the pure state
$\ket\psi=\ket{\psi(\eta,\theta)}$ given by
\begin{equation}
\ket \psi  =\left(
\begin{array}{c}
e^{-i\theta/2}\cos(\eta/2) \\
e^{i\theta/2}\sin(\eta/2)
\end{array}
\right) .
\label{example}
\end{equation}
This pure state has density matrix
$\rho= =
\half(\mathbf 1+\vec u\cdot\vec\sigma)$
where $\vec u=\vec u(\eta,\theta)$
is the point on the unit sphere
in $\openR^3$ with polar coordinates $(\eta,\theta)$.
Suppose the colatitude $\eta\in[0,\pi]$ is known
and exclude the degenerate cases $\eta=0$ or $\eta=\pi$; the
longitude $\theta \in [0,2\pi)$ is the unknown parameter.

We have a pure state so $\lambda= 2\dot\rho=2\dot{\vec u}
\cdot\vec\sigma=\sin(\eta)\,\vec u(\pi/2,\theta+\pi/2)\cdot\vec\sigma
=r(\theta)\vec v(\theta)$.
The quantum information is $r^2=\sin^2\eta$. As $\theta$ varies,
$\vec u(\theta)$ traces out a great circle if and only if $\eta=\pi/2$.
Consequently, for $\eta\ne \pi/2$, no measurement $M$
exists with Fisher information $i(\theta;M)$ equal to the
quantum information $I(\theta)$ whatever the value of
the unknown parameter $\theta$.
If $\eta=\pi/2$ it {\em is\/} possible
to achieve the bound uniformly in $\theta$.
Any measurement with everywhere positive density and
all components proportional to projector matrices for
spin directions in the plane $\eta=\pi/2$ will do the job.
A simple measurement of spin in one particular direction
in that plane attains the information bound at all $\theta$
except for $\theta$ equal to that direction or opposite to it.
At these points the distribution of the outcome is
degenerate and the Fisher information not defined. However
since the Fisher information is continuous (indeed, constant)
in $\theta$ this is a non-essential singularity.

\section{Asymptotic attainability and vector parameters}

We have shown, for the case of a one-dimensional parameter,
that only for rather special models will a measurement $M$ exist such
that $i(\theta;M)=I(\theta)$ for all parameter values $\theta$
simultaneously. It is on the other hand possible to find a
measurement $M$ such that at a {\em given\/}
parameter-value, $i(\theta;M)=I(\theta)$, as
Braunstein and Caves indicate: take each $m(x)$ proportional
to a projector onto an eigenspace of the quantum score $\lambda(\theta)$.
They do not remark on the possible dependence of $M$ on $\theta$.
However, if all we know is that $\rho=\rho(\theta)$ for {\em some\/} $\theta$,
we do not know which measurement to use.
The eigenspace decomposition of $\lambda$ generally depends on $\theta$
so this does not define a measurent $M$ which achieves the bound uniformly
in $\theta$. This is not the only way to achieve the bound,
but the previous section shows that one cannot in general expect
there to be a uniformly attaining measurement.

Note that the classical information based on $n$ independent and
identically distributed realisations from a given density $p(x,\theta)$
is equal to $n$ times the information for one realisation.
Similarly, the quantum information in the state $\rho(\theta)^{\otimes n}$
corresponding to $n$ identical particles each in state $\rho(\theta)$
is $n$ times the quantum information for one particle.

Braunstein and Caves' aim was to define a statistical
distinguishability metric between quantum states.
Suppose the measurement $M$ on a single particle satisfies
$i(\theta;M)=I(\theta)$. Then
the maximum likelihood estimator of $\theta$
based on $n$ separate measurements of $M$ on identical
copies of the given quantum system, by classical results in
mathematical statistics, is generally an asymptotically unbiased estimator
with {\it asymptotic\/} variance $(n i(\theta;M))^{-1}=(n I(\theta))^{-1}$.
By the quantum Cram\'er-Rao bound applied to the joint system
of $n$ particles, no estimator based on any measurement whatsoever
on $\rho^{\otimes n}(\theta)$ can do better. Thus $I(\theta)$ appears
to exactly characterize the rate at which one can determine $\theta$.

However this argument is flawed since the measurement $M$ involved
will be a different measurement for each $\theta$, and the whole
point is that $\theta$ is not known in advance.
The question therefore remains:
does there exist a measurement procedure {\em not depending on
$\theta$\/}
on the state  $\rho^{\otimes n}$, on the basis of which an
estimator of $\theta$ can be constructed having asymptotic
variance $(nI(\theta))^{-1}$? If the answer is `yes',
then Braunstein and Caves' proposed role
for the quantum information $I(\theta)$
in defining a statistical distinguishability metric is well motivated.

It seems rather natural to try a two-stage procedure:
first estimate the parameter using a perhaps inefficient procedure
on a vanishing proportion of the particles,
say $n_0=n^\alpha$ ($0<\alpha<1$) out of the total of $n$;
now carry out the `estimated optimal measurement' on the remaining
ones. In both stages only simple or von Neumann measurements
(measurements of classical observables) on separate particles are needed.

In our example \eqref{example} this would reduce to the following.
Measure the spin $\sigma_x$ on $k=\half n_0$ of the copies.
The number of $+1$'s observed is binomially distributed with parameters
$k$ and $p=\half(\sin\eta\cos\theta+1)$.
Similarly for another $k$  measurements of the spin $\sigma_y$
we get a binomial number of `$+1$' with parameters $k$ and
$p=\frac{1}{2}(\sin\eta\sin\theta+1)$. This allows us consistent
estimation of both $\sin\theta$ and $\cos\theta$
and hence of $\theta\in[0,2\pi)$.
Denote such an estimator by $\widetilde\theta$.
We saw that $\lambda$ in this example was proportional to
the spin in the direction $(\pi/2,\theta+\pi/2)$. Let us use
the remaining $n'=n-n_0$ particles to measure this spin with
$\theta$ replaced by $\widetilde \theta$. Given $\widetilde\theta$,
this results in a binomial number $X$ of `$+1$' with parameters
$n'$ and $p=\half(1-\sin\eta\sin(\theta-\widetilde\theta))$.
Let
\begin{equation*}
\widehat\theta=\widetilde\theta+\arcsin((n'-2X)/(n'\sin\eta)) .
\end{equation*}
Analysis of this `final' estimator
shows that $\widehat\theta$ has asymptotically the
${\mathcal N} (\theta,(n\sin^2(\eta))^{-1})$
distribution (the normal distribution with indicated mean and variance),
whatever $\theta$, so that the quantum information bound
is asymptotically achievable by our two stage procedure.

This approach will work in wide
generality in problems with a one-dimensional parameter $\theta$.
Suppose, as typically will be the case, that one can construct a
consistent estimator $\widetilde\theta$ based on certain measurements
on a vanishing proportion of the particles.
Compute the quantum score at $\theta=\widetilde\theta$, and measure it
on each of the remaining particles. Compute the maximum likelihood estimator
$\widehat\theta$ of $\theta$ based on the new data,
whose probability distribution depends
on the unknown $\theta$ (as well as on $\widetilde\theta$, which is
at this stage fixed).
We argue as follows that $\widehat\theta$ has approximately the
${\mathcal N}(\theta,(nI(\theta))^{-1})$ distribution, thus this estimator
asymptotically achieves the quantum information bound.
Let $i(\theta;\widetilde\theta)$
denote the Fisher information for $\theta$ in a measurement,
on one particle, of the quantum score at $\widetilde\theta$;
thus $i(\widetilde\theta;\widetilde\theta)=I(\widetilde\theta)$
for all values of $\widetilde\theta$, but generally
$i(\theta;\widetilde\theta)<I(\theta)$. Now for $n$ large,
$\widetilde\theta$
is close to $\theta$. By the classical results for maximum likelihood
estimators,
given $\widetilde\theta$, $\widehat\theta$ has approximately the
${\mathcal N}(\theta,(ni(\theta;\widetilde\theta))^{-1})$ distribution.
So if $\rho$ depends on $\theta$ smoothly enough that
$i(\theta;\widetilde\theta)$ is close to
$i(\theta;\theta)=I(\theta)$ for $\widetilde\theta$ close to $\theta$,
we have that unconditionally $\widehat\theta$ has approximately the
${\mathcal N}(\theta,(nI(\theta))^{-1})$ distribution, hence asymptotically
achieves the bound.

Consider now the case of vector parameters.
Both quantum and Fisher information numbers are naturally
generalized to information matrices; see Helstrom (1976), Holevo (1982).
The Braunstein and Caves result
generalizes to the following result: the quantum information matrix
is larger (in the sense that the difference is positive semi-definite)
than the Fisher information matrix based on the outcome of
any measurement $M$. However the bound is no longer attainable.
As we saw above, a best measurement for each parameter separately
is a measurement of the quantum score operator. Typically
these do not commute and hence cannot be measured simultaneously.
On the other hand, consideration of all
smooth one-dimensional sub-models of a given
model $\rho=\rho(\theta)$ shows that the quantum information
matrix {\em\/} is the smallest matrix larger than the Fisher information
matrix of any measurement on a single particle.

For instance, suppose we want to simultaneously estimate
both parameters $\eta$, $\theta$ of the pure-state,
spin-half system \eqref{example};
in other words, we have a completely unknown pure state.
Rename $\theta$ as $\phi$, and let $\theta$ from now
denote the vector parameter with elements $\eta$, $\phi$.
Suppose we may dispose of a large number of identical copies of this system.
Let $I(\theta)$ denote the $2\times 2$ quantum
information matrix, and $i(\theta;M)$ denote the Fisher information matrix
based on the outcome of a measurement $M$, both for a single copy of the
quantum system. The quantum scores, for a single particle,
for the two parameters $\eta$ and $\phi$ are $\sigma_{\eta+\pi/2,\phi}$
and  $\sin\eta\sigma_{\pi/2,\phi+\pi/2}$ respectively.
After a small proportion of measurements we know roughly the
location of the parameter, and it is sufficient to investigate optimal
measurement at a `known' parameter value.

Without loss of generality let this be the special point $\theta_0$
with $\eta_0=\pi/2$, $\phi_0=0$; $\rho=\half(\mathbf 1 +\sigma_x)$.
At this point the quantum scores
are $\sigma_y$ and $-\sigma_z$, and the quantum information matrix is the
identity $\bf 1$. The intersection of the $xy$ and the $xz$ planes is
the $x$ axis, so it seems that in order to maximize Fisher information
both for $\eta$ and for $\phi$ we should essentially measure spin in the $x$
direction. However the Fisher information matrix $i(\theta;M)$ based on
the outcome of this measurement is not defined at
$\theta=(\pi/2,0)=\theta_0$ since the
distribution of the outcome is degenerate. The singularity is
essential since $i(\theta;M)$ is not continuous at $\theta=\theta_0$.
In particular, as one moves towards $\theta_0$ along either of the two
great circles formed by varying one of the two components of
$\theta$, $i(\theta;M)$ converges to each of the diagonal matrices with
diagonal elements $1,0$ and $0,1$. These are the two Fisher information matrices
corresponding to von Neumann measurements of the two quantum scores,
each giving maximal information about the corresponding component
of $\theta$ and zero information about the other.

Since different components of the parameter vector have incompatible
quantum scores, it is clear that for different loss functions,
different measurements will be optimal. No single procedure will
(asymptotically) dominate all others. Moreover, since we cannot
achieve the quantum information bound by measurements on single particles,
it is possible that joint measurements
on several particles simultaneously could give larger Fisher information
(per particle) than measurements on separate particles.

In some very special cases, an optimal procedure is known.
An appealing loss function in the completely unknown pure spin-half model
is one minus the squared inner-product between the true state vector and
its estimate. This equals one minus the squared cosine of half the angle
between the points on the Poincar\'e sphere representing the two states. 
At the special point under consideration therefore, the loss function is
asymptotically equivalent to one quarter times the sum of the
squares of the errors in $\eta$ and $\phi$.
Massar and Popescu (1995), in response to a problem
posed by Peres and Wootters (1991), exhibited a measurement, optimal
in the Bayes sense, with respect to this loss function
and a uniform prior distribution. It had an asymptotic mean square error $4/n$.
This was a genuine generalised measurement of the composite
system $\rho^{\otimes n}$. They showed that for the case of $n=2$
there were no measurement methods of the two particles separately
which were as good as the optimal method, and this is expected to hold
for all $n$.

Instead of this exactly optimal procedure (with respect to the given
loss function and under a uniform prior)
consider taking with probability half measurements of
$\sigma_y$ and $\sigma_z$, independently on each particle.
We find that the Fisher information matrix (based on one observation)
for $\eta,\phi$, at $\eta=\pi/2$, $\phi=0$, is $\frac 12 {\bf 1}$,
or one half of the quantum information matrix.
The inverse of this matrix, $2\,{\bf 1}/n$ is an asymptotically
achievable lower bound to the covariance matrix
of (asymptotically unbiased) estimators of $\eta,\phi$
based on $n$ of such measurements. The maximum likelihood method
would provide estimators asymptotically achieving this bound.
The sum of the variances is $4/n$, the same as what is
achieved by the Massar and Popescu procedure.

Thus the following two-stage procedure, similar
to what we proposed in the one-parameter case, should have asymptotically
equivalent covariance matrix to that of the Massar and Popescu procedure,
and will also be optimal with respect to a uniform prior distribution and
any smooth loss function, invariant under rotations of the sphere.
First carry out measurements of each of $\sigma_x$, $\sigma_y$ and $\sigma_z$
on a small proportion of separate particles.
Compute from the results a consistent
estimate of $(\eta,\phi)$. With respect to a rotated coordinate system
putting the estimated value at $(\eta,\phi)=(\pi/2,0)$,
measure alternately $\sigma_y$ and $\sigma_z$ on the
remaining particles. Estimate
$(\eta, \phi)$ in the new coordinate system by the method of
maximum likelihood using the second stage observations. Finally
rotate back to the original coordinate system.
Note that if we modified this scheme by just
measuring $\sigma_x$ in the second stage, then overall the procedure
determines the radial distance of $\theta$
from $\tilde\theta$ with precision of the order $n^{-1/2}$, but says nothing
about the direction, so that finally $\theta$ has not been localized
to this precision at all.

This conjecture has been confirmed by recent further
work of Gill and Massar (1999).
Consider a sequence of measurements on
$n$ identical copies of a spin-half state,
on which is based a sequence of estimators $\hat\theta$. The parameter
$\theta$ might be one- or two-dimensional for a pure state model, one-,
two- or three-dimensional for a mixed state. Suppose that the
estimators are asymptotically unbiased
and have covariance matrices asymptotically
of the form $V(\theta)/n$. Then it is now known
that the collection of attainable
$V$ is precisely $\{V:\mbox{trace}\, I(\theta)^{-1} V(\theta)^{-1}\le 1\ 
\hbox{for all}\ \theta\}$,
in any of the following cases: the parameter is one-dimensional, or the
state is pure, or the measurement can be implemented by separate measurements
on separate particles.
These optimal limiting covariance matrices can all be achieved using a
two stage adaptive procedure of the type described above.
The collection of attainable $V$ corresponds
to the collection of attainable inverse Fisher information matrices for
measurements on single particles.
Using joint measurements on mixed states with more than one unknown
parameter, one can attain strictly smaller asymptotic covariance matrices.
But a clean description of what is attainable is not known.
One would like to describe the collections of
scaled information matrices $\{i(\theta;M_m)/m\}$ for $m=1,2,\dots$,
where $M_m$ is an arbitrary joint measurement on $m$ particles. These
sets are all convex, they grow with $m$; we know the set for $m=1$; and
each set is included in the set of matrices less than or equal to
$I(\theta)$. The inverses of these information matrices
will be the achievable (scaled) asymptotic covariance matrices based on
measuring a large number $n$ of particles in groups of $m$ at a time.

To conclude, in the multiparameter case, the
bound implied by the quantum information matrix is
not even asymptotically achievable. The rate at which one can
distinguish between more than two neighbouring quantum states
does not correpond to the rate at which one can distinguish between
just two; it depends on what aspect of the quantum states is
important, and it depends on whether one may use joint measurements
or only separate measurements. The quantum information matrix only
plays a role in special cases.

\section{Acknowledgements}

We are grateful for useful discussions with
Alessandra Luati, Klaus M\o lmer, and Peter Jupp.
This research was supported by the Danish National
Research Foundation through MaPhySto (Centre for Mathematical Physics
and Stochastics), and by the Oberwolfach Mathematical Institute's Research
in Pairs programme. Richard Gill is also grateful for the hospitality of
the Department of Mathematics and Statistics, University of Western Australia.

\section*{References}

\begin{harvard}

\item[]
Barndorff-Nielsen O E and Gill R D
1998
An example of non-attainability of expected quantum information
{\tt xxx.lanl.gov, quant-ph/9808009}

\item[]
Belavkin V P
1976
Generalized uncertainty relations and efficient measurements in
quantum systems
{\it Teoreticheskaya i Matematicheskaya Fizika}
{\bf 26}
316--329
English translation, 213--222.

\item[]
Braunstein S L and Caves C M
1994
Statistical distance and the geometry of quantum states
{\it Phys.\ Review Letters}
{\bf 72}
3439--3443

\item[]
Gill R D and Massar S
1999
State estimation for large ensembles
{\tt xxx.lanl.gov, quant-ph/9902063}; {\it Phys.\ Rev. A} (to appear)

\item[]
Helstrom C W
1967
Minimum mean-square error of estimates in quantum statistics
{\it Phys.\ Letters}
{\bf 25 A}
101--102

\item[]
Helstrom C W
1976
{\it Quantum Detection Theory}
(New York: Academic)

\item[]
Holevo A S
1982
{\it Probabilistic and Statistical Aspects of Quantum Theory}
(Amsterdam: North-Holland)
Russian original, 1980

\item[]
Massar S and Popescu S
1995
Optimal extraction of information from finite quantum ensembles
{\it Phys.\ Review Letters}
{\bf 74}
1259--1263

\item[]
Peres A and Wootters W K
1991
Optimal detection of quantum information
{\it Phys.\ Review Letters},
{\bf 66}
1119--1122

\item[]
Yuen H P and Lax M
1973
Multiple-parameter quantum estimation and measurement of
nonselfadjiont observables
{\it Trans.\ IEEE}
{\bf IT-19}
740--750

\end{harvard}

\end{document}